\documentclass{revtex4}
\usepackage{graphicx}
\usepackage{fancyhdr}
\usepackage{amsmath}
\def\bma{\left( \begin{array} }
\def\ema{\end{array} \right)}
\newcommand{\vdel}{v_{\Delta}}
\newcommand{\delm}{\Delta M}

\usepackage{color}

\begin{document}

\title{Same-sign tetra-leptons in type II seesaw at the LHC}

%

\author{Eung Jin Chun and Pankaj Sharma}
\affiliation{Korea Institute for Advanced Study, Seoul 130-722, Korea}

\begin{abstract}
In type II seesaw model of neutrino mass generation, we study a remarkable 
signal of same-sign tetra-lepton (SS4L) signal at the Large hadron collider. When 
doubly charged Higgs is lightest, the heavier singly charged or neutral Higgs boson
produces a doubly charged Higgs boson through its fast gauge
decay. This leads to a novel signature of same-sign tetra-leptons
resulting from a pair production of same-sign doubly charged Higgs
bosons. We study production cross section for the SS4L signal in parameter 
space of the mass splitting among triplet components and
 the triplet vacuum expectation value at the LHC.
\end{abstract}

\maketitle

\thispagestyle{fancy}


\section{Introduction}
One of the important question in particle physics is the origin of neutrino masses and mixing. In 
type II seesaw model  \cite{magg}, Higgs triplet coupling to lepton doublet can generate neutrino mass terms once 
it develops non-trivial vacuum expectation value. A distinct feature of this model is the presence of 
doubly charged Higgs bosons which can give a very clean signal at collider experiments 
\cite{gunion}--\cite{Muhlleitner:2003me}. Moreover it can 
also explain the deviation in Higgs-to-diphoton rate as measured by CMS and ATLAS \cite{h2gg}.

In this work \cite{SS4L}, we study a novel signature of same-sign tetra-leptons (SS4L) allowed in some parameter space of the type II
seesaw model that has not been studied so far. When
the mass splitting among triplet components is sizable and the doubly charged Higgs boson
is the lightest, the electroweak gauge interaction allows a fast
decay of the neutral or singly charged component of the Higgs
triplet into the lighter singly or doubly charged component.
Therefore, pair-produced Higgs triplet components can end up with
a pair of same-sign doubly charged Higgs bosons leading to
same-sign tetra-leptons if their leptonic Yukawa coupling is larger
than the ratio of the triplet and doublet Higgs
vacuum expectation values.

The SS4L final state,
which is almost background free, provides an excellent new channel
to test the model and probe sizes of the Higgs triplet vacuum
expectation value and the mass splitting among the Higgs triplet
components at the LHC.
%

\section{The Type II Seesaw Model}

In this model, the Higgs sector of the Standard Model is extended with a
$Y=2$ $SU(2)_L$ scalar triplet $\Delta$ in addition to a
SM-Higgs doublet $\Phi$. 
The leptonic part of the Lagrangian which generates
neutrino masses is
\begin{equation} \label{leptonYuk}
\mathcal L_Y= f_{\alpha\beta}L_\alpha^T Ci\tau_2\Delta L_{\beta} +
\mbox{H.c.},
\end{equation}
and the scalar potential is
\begin{eqnarray}\label{Pot}
V(\Phi,\Delta)&=&\nonumber m^2\Phi^\dagger\Phi +
\lambda_1(\Phi^\dagger\Phi)^2+M^2\mbox{Tr}(\Delta^\dagger\Delta)
+\lambda_2\left[\mbox{Tr}(\Delta^\dagger\Delta)\right]^2+\lambda_3\mbox{Det}(\Delta^\dagger\Delta)
+\lambda_4(\Phi^\dagger\Phi)\mbox{Tr}(\Delta^\dagger\Delta)\\
&+&\lambda_5(\Phi^\dagger\tau_i\Phi)\mbox{Tr}(\Delta^\dagger\tau_i\Delta)
+\left[\frac{1}{\sqrt{2}}\mu(\Phi^Ti\tau_2\Delta\Phi)+\mbox{H.c.}\right].
\end{eqnarray}

After the electroweak symmetry breaking with $\langle
\Phi^0\rangle=v_0/\sqrt{2}$, the $\mu$ term in Eq.~(\ref{Pot}) gives rise to the
vacuum expectation value of the triplet $\langle
\Delta^0\rangle=v_\Delta/\sqrt{2}$ where $\vdel\approx \mu
v_0^2/\sqrt{2}M^2$ with $\mu$ being real and positive.

There are seven physical
massive scalar eigenstates denoted by $H^{\pm,\pm}$, $H^\pm$,
$H^0$, $A^0$, $h^0$. Under the condition that $|\xi|\ll 1$ where
$\xi \equiv v_\Delta/v_0$, the mixing among doublet and triplet components is too small and thus
the first five states are mainly from the triplet and the last from the doublet. 
We write the masses of the Higgs bosons as
\begin{eqnarray} \label{massD}
 M^2_{H^{\pm\pm}} &=& M^2 + 2{\lambda_4 -\lambda_5 \over g^2 } M^2_{W}
 \nonumber\\
 M^2_{H^{\pm}} &=& M_{H^{\pm\pm}}^2 + 2{\lambda_5 \over g^2} M^2_{W}
 \nonumber\\
 M^2_{H^0, A^0} &=&  M^2_{H^\pm} +
    2{\lambda_5 \over g^2} M^2_{W} \,.
\end{eqnarray}
The mass of $h^0$ is given by $m_{h^0}^2=4\lambda_1 v_\Phi^2$ as usual.

The mass splitting among triplet scalars can be approximated as
\begin{equation}
 \delm \approx \frac{\lambda_5 M_W^2}{g^2 M} < M_W \,.
\end{equation}
Furthermore, the sign of the coupling $\lambda_5$
determines two mass hierarchies among the triplet components:
$M_{H^{\pm\pm}}>M_{H^\pm}>M_{H^0/A^0}$ for $\lambda_5<0$; or
$M_{H^{\pm\pm}}<M_{H^\pm}<M_{H^0/A^0}$ for $\lambda_5> 0$. In this
work, we focus on the scenario, where the doubly charged
scalar $H^{\pm\pm}$ is the lightest and thus it decays only to
$l_\alpha^\pm l_\beta^{\pm}$ or $W^\pm W^\pm$, 
while $H^0/A^0$  ($H^\pm$) decays
mainly to $H^\pm W^{\mp *}$ ($H^{\pm\pm} W^{\mp *}$) unless the
mass splitting $\Delta M$ is negligibly small. For more details,
see, e.g., Ref.~\cite{Chun:2003ej}.

The mass splitting $\delta M_{HA}$ between $H^0$ and $A^0$ plays a crucial role for 
SS4L signal. The $\mu$ term in Eq.~(\ref{Pot}),
which is lepton number violating, generates not only the triplet VEV $v_\Delta$ but also
$\delta M_{HA} \equiv M_{H^0} - M_{A^0}$:
\begin{equation} \label{MHA}
 \delta M_{HA} = 2 M_{H^0} {v_\Delta^2\over v_0^2}
 { M^2_{H^0} \over M^2_{H^0} - m^2_{h^0} } \,.
\end{equation}
For a preferable choice of $v_\Delta$, $\delta M_{HA}$ can be comparable to the total decay rate
of the neutral scalars, $\Gamma_{H^0/A^0}$, enhancing the SS4L signal.


\section{Decays of triplet scalars}

Depending on the triplet mass splitting $\Delta M$ and the triplet vacuum
expectation value $v_\Delta$, the triplet scalars have different decay properties.
When $H^{++}$ is the lightest, the possible decay channels for the
triplet scalars are shown in Table.~\ref{decay}. The decays of triplet scalars can be classified into three modes,
a) di-leptonic mode: which is proportional to Yukawa coupling $f$, b) quark/di-boson mode: controlled by $\xi=v_\Delta/v_0$,
and c) gauge decay: due to $SU(2)$ gauge interaction which dominates if allowed
kinematically.
 

\begin{table}[h]
\begin{center}
\begin{tabular}{|l|l|l|l|}
\hline $~~~~~~H^0$ &   $~~~~~~A^0$ & $~~~~~~H^+$ & $~~~~~H^{++}$
\\\hline
$\to t\bar t/b\bar b$  &   $\to t\bar t/b\bar b$           &   $\to t\bar b$       &   $\to \ell^+ \ell^+ $\\
$\to \nu\bar{\nu}$     &   $\to \nu\bar{\nu}$           &   $\to \ell^+\nu$     &   $\to W^{+} W^{+} $\\
$\to ZZ$  		&   $\to Zh^0$      &   $\to W^+Z$      &   \\
$\to h^0h^0 $       	&   $\to H^{\pm}W^{\mp^*}$          &   $\to W^+h^0$        &   \\
$\to H^{\pm}W^{\mp^*}$ &         &   $\to H^{++}W^{-^*}$ &   \\\hline
\end{tabular}\caption{\label{decay}Possible decay channels for the triplet Higgs bosons for $\lambda_5>0$.}
\end{center}
\end{table}

\begin{figure}[h]
\begin{center}
\includegraphics[width=55mm]{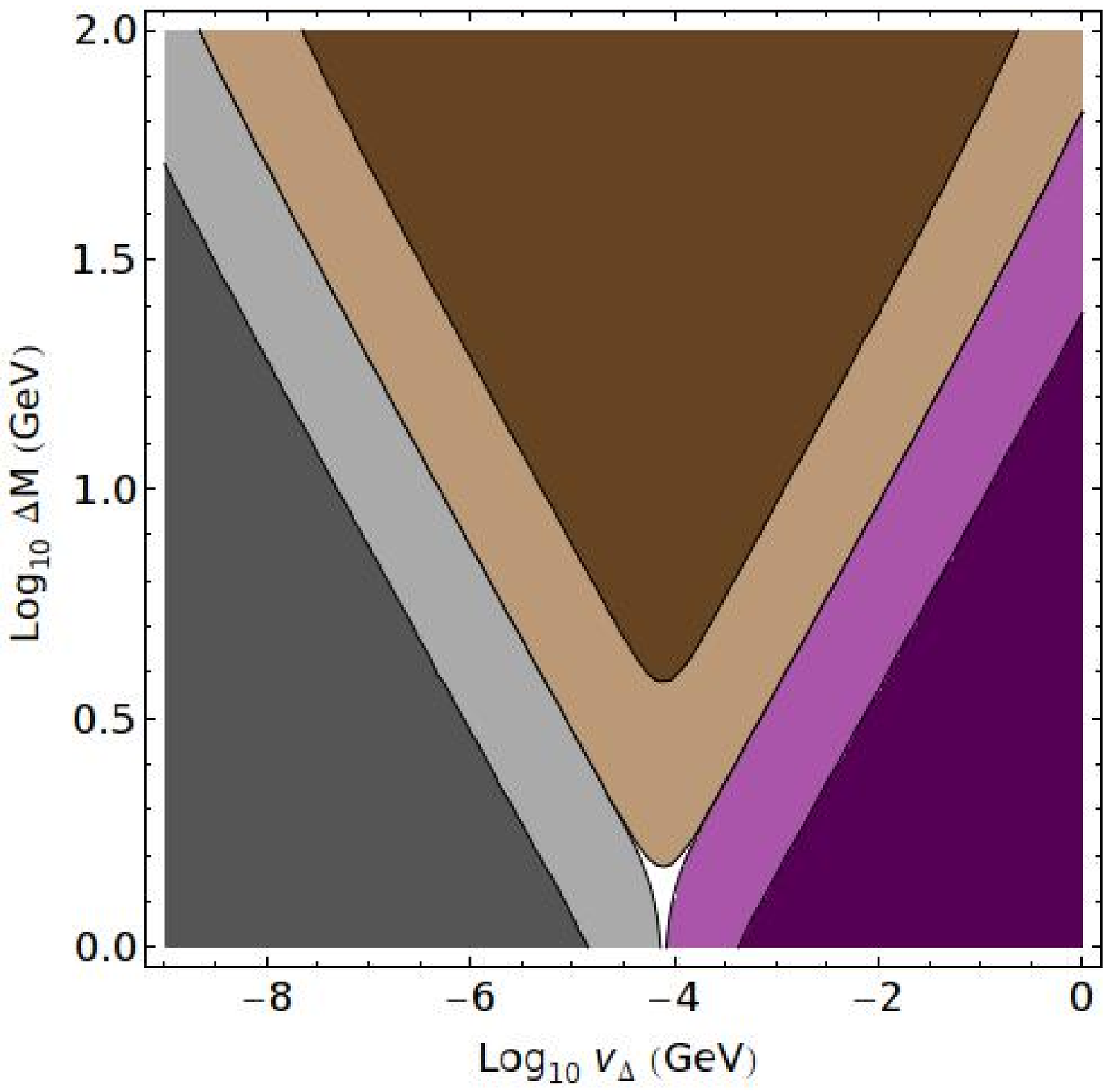}
\includegraphics[width=55mm]{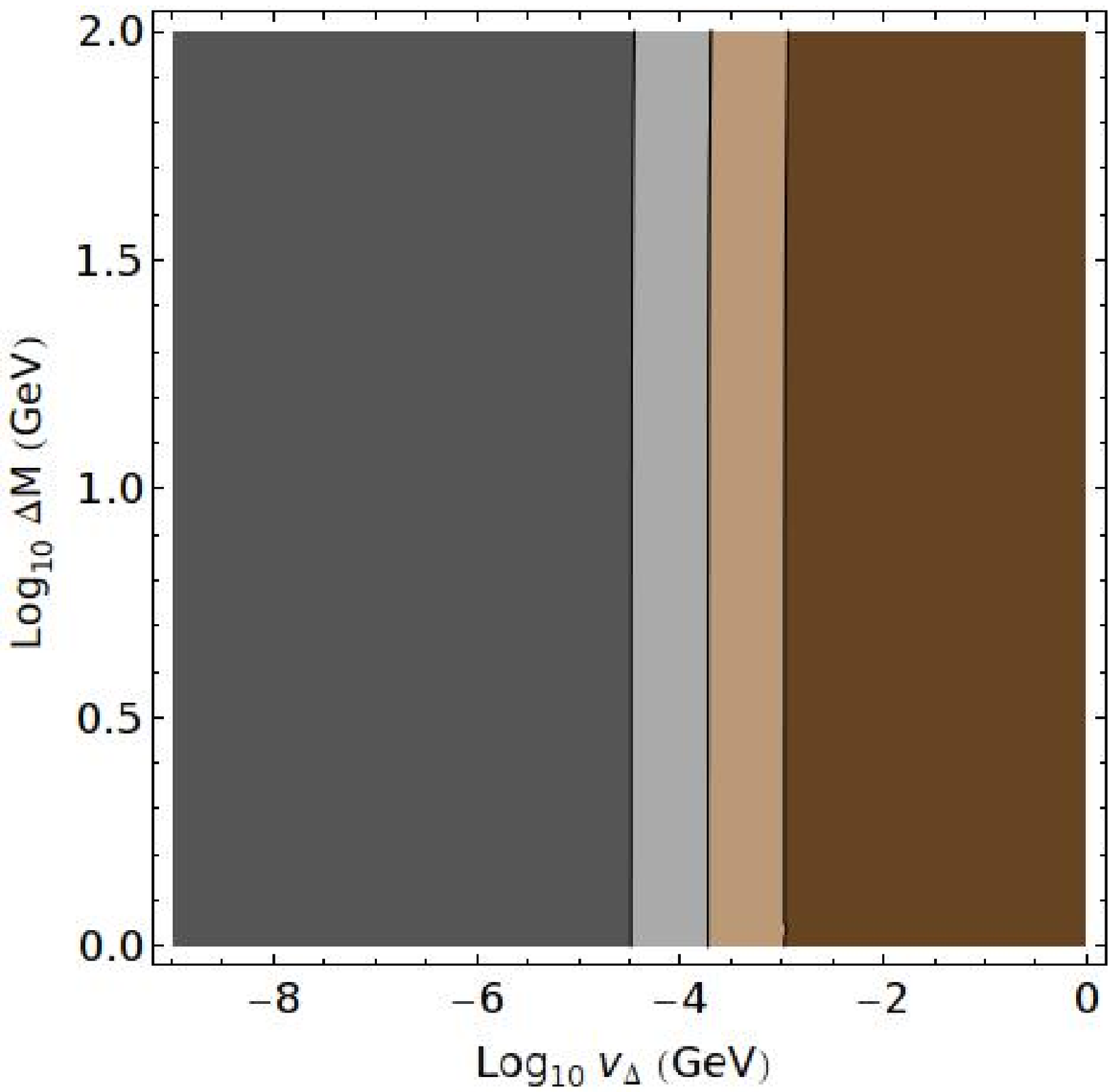}
\end{center}
\caption{ Phase diagrams for $H^+$(left) and $H^{++}$(right)
decays in the type II seesaw model for $\lambda_5>0$ with
$M_H^{++}=300$ GeV. The dark-colored regions denote the branching
fraction larger than 99\%.}\label{decay:Hp}
\end{figure}

In Fig.~\ref{decay:Hp}, we
show phase diagrams for $H^+$ and $H^{++}$ decays in the plane of
$\Delta M$ and $\vdel$. In the left panel, the brown, the gray
and the purple regions show the branching fractions for the decays
$H^+\to H^{++}W^{-^*}$, $H^+\to\ell^+\nu$ and $H^+\to \{t\bar
b,W^+Z,W^+h\},$ respectively. In the right panel, the brown and
the gray regions show the branching fractions for the decays
$H^{++}\to W^+W^+$ and $H^{++}\to \ell^+\ell^+$ respectively. In
both panels, the dark-colored regions denote the parameter space
where the branching fraction is greater than 99\% and the
light-colored regions denote the parameter space where the
branching fraction is between 50\%-99\%.

As it can be seen from Fig.~\ref{decay:Hp}, the leptonic decay BF of
$H^+$ is dominant for small values of $v_\Delta<0.1$ MeV and always greater
than 0.99. However, for moderate mass splitting $\delm>5$ GeV, the
gauge decay of $H^+$ starts becoming large at $v_\Delta=0.1$ MeV and at
$\Delta M$ around 10-20 GeV, large parameter space opens up
for this decay. Hence, for moderate $v_\Delta$
(around 1 keV-10 MeV), large mass splitting is allowed for the
BF larger than 99\%. On the other hand, for large
$v_\Delta$ and low mass splitting, rest of the decays viz.,
$(H^+\to t\bar b, W^+Z, W^+ h)$ have appreciable contributions and
thus BF($H^+\to H^{++}W^{-^*})$ goes down.

In the case $H^{\pm\pm}$, for $v_\Delta<0.1$ MeV,
it is completely dominated by leptonic decay i.e., $H^{\pm\pm}\to
\ell^\pm\ell^\pm$ while for $v_\Delta>1$ MeV, it is dominated by
the di-$W$ decay. These BFs are completely independent of
the mass splitting $\Delta M$ as the $H^{\pm\pm}$ is the lightest.

\begin{figure}[h]
\begin{center}
\includegraphics[width=55mm]{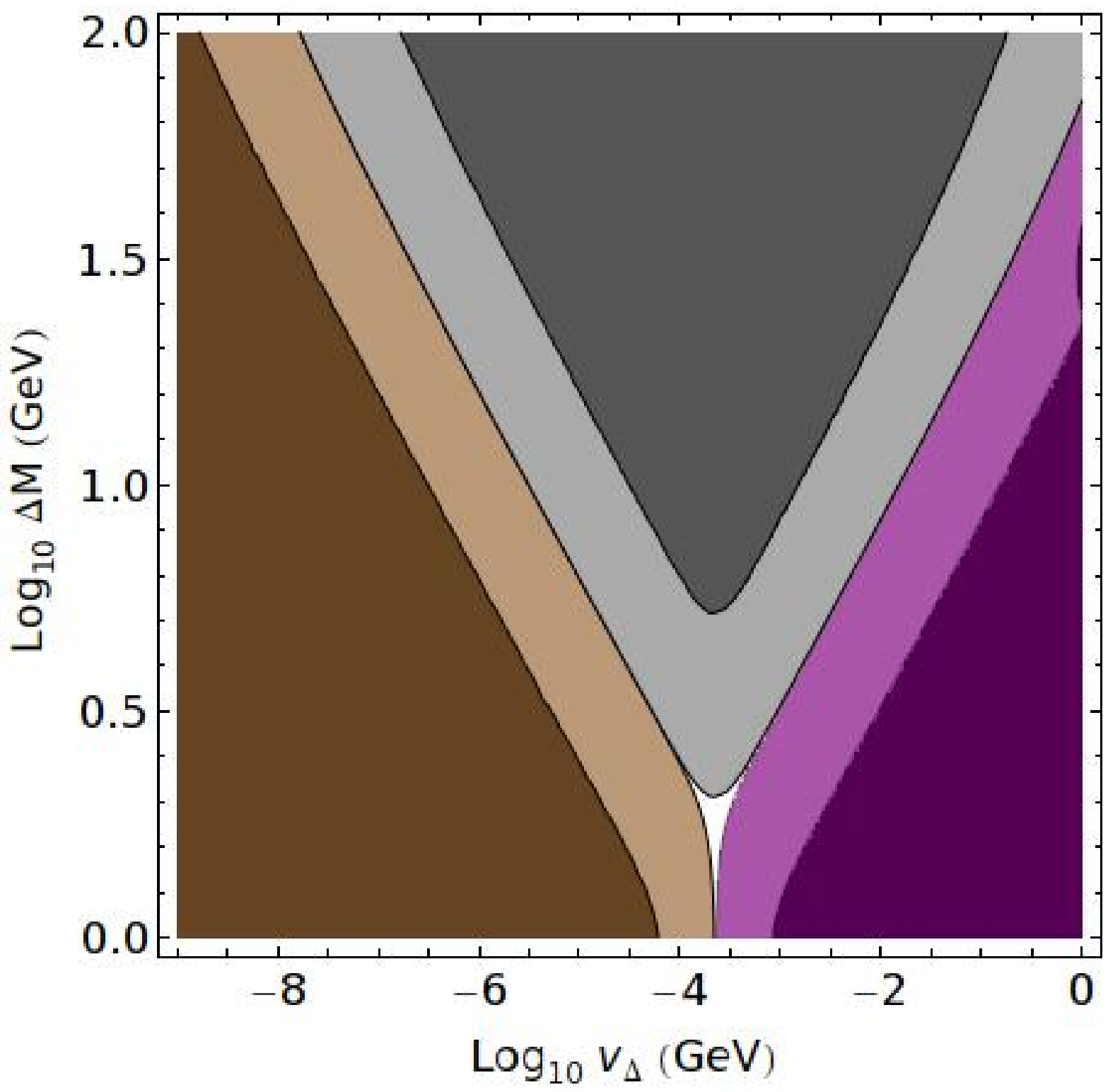}
\includegraphics[width=55mm]{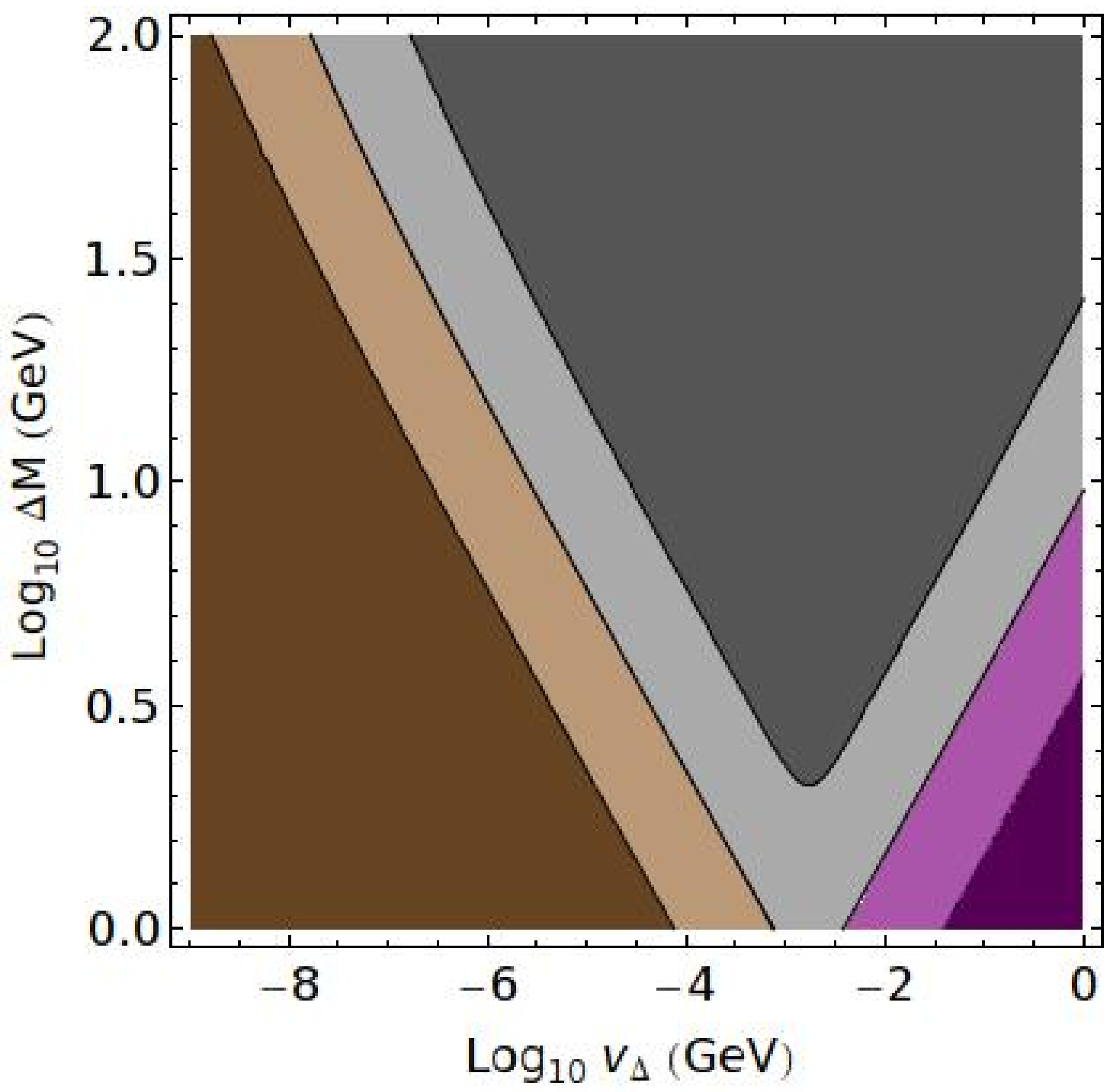}
\end{center}
\caption{ Phase diagrams for $H^0$(left) and $A^0$(right)  decays
in the type II seesaw model for $\lambda_5>0$ with $M_H^{++}=300$
GeV. The dark-colored regions denote the branching fraction in the
range of 49\%-50\%.}\label{decay:H0}
\end{figure}

Similarly, in Fig.~\ref{decay:H0}, we show phase diagrams for
$H^0$ and $A^0$ decays in the plane of $\delm$ and $\vdel$. In the
left panel, the brown, the gray and the purple regions show the
BFs for the decays $H^0\to\nu\bar{\nu}$, $H^0\to H^{+}W^{-^*}$,
and $H^0\to \{t\bar t, b\bar b, ZZ,h^0h^0\}$ respectively. In the
right panel, the brown, the gray and the purple regions show the
BFs for the decays $A^0\to\nu\bar{\nu}$, $A^0\to H^{+}W^{-^*}$,
and $A^0\to \{t\bar t, b\bar b, Zh^0\}$ respectively. In both
panels, the dark-colored regions denote the parameter space where
BF is between 49\%-50\% and the light-colored regions denote the
parameter space where the BF is between 20\%-49\%.
\section{Same-Sign Tetra-Leptons at the LHC}

In this section, we study a new possibility of
probing the type II seesaw model at the LHC through a remarkable
signal of four same-sign leptons which are either positively or
negatively charged.  The processes which contribute to such a
signal are as follows:
\begin{enumerate}
\item\label{pro:1} $q^\prime \bar q\to W^*\to H^{\pm}H^0/A^0$
proceeded by $H^\pm\to H^{\pm\pm}W^{\mp^*}$  and $H^0/A^0\to
H^{\pm}W^{\mp^*}\to H^{\pm\pm}W^{\mp^*}W^{\mp^*}$;
\item\label{pro:3} $q \bar q\to Z^*\to H^0A^0$
proceeded by  $H^0/A^0\to H^{\pm}W^{\mp^*}\to
H^{\pm\pm}W^{\mp^*}W^{\mp^*}$.
\end{enumerate}
Out of all the triplet
pairs produced in the above three processes, only some fraction of them
eventually give same-sign $H^{\pm\pm}$ pairs which is controlled
by the neutral scalar mixing parameter as will be discussed below.


In the limit of lepton number conservation i.e. $\mu, v_\Delta \to 0$, the SS4L final state cannot occur,
due to the canceling interference between $H^0$ and $A^0$ ~\cite{Akeroyd:2011zz}.
However, the above processes 1 and 2 are allowed when there is a finite mass difference $\delta M_{HA}$
(\ref{MHA}) violating lepton number.
Like as in the $B^0$-$\bar{B}^0$ system, this phenomenon occurs as $H^0$ and $A^0$,
sharing the same final states, can mix together .
If they undergo sufficient oscillation before they decay, i.e., $\delta M_{HA} \gtrsim \Gamma_{H^0/A^0}$,
the lepton number violating production of SS4L becomes sizable.  This effect is controlled by the usual oscillation parameter $x_{HA}$:
\begin{equation}
 x_{HA} \equiv { \delta\! M_{HA} \over \Gamma_{H^0/A^0} } \,.
\end{equation}
Using the narrow width approximation for the calculation of the $H^0$-$A^0$ interference term, we obtain the cross-sections for the processes \ref{pro:1} and \ref{pro:3} as follows:
\begin{eqnarray}
\sigma\left(4\ell^\pm + 3W^{\mp^*}\right)&=& \sigma\left(pp\to
H^\pm H^0 + H^\pm A^0\right)  \left[{ x_{HA}^2\over 1+ x_{HA}^2}\right]
 \mbox{BF}(H^0/A^0\to H^\pm W^{\mp^*}) \nonumber\\
&\times& \left[\mbox{BF}(H^\pm\to H^{\pm\pm} W^{\mp^*})\right]^2
\left[\mbox{BF}(H^{\pm\pm}\to \ell^\pm\ell^\pm)\right]^2;
\label{4l3W}\\
\sigma\left(4\ell^\pm + 4W^{\mp^*}\right)&=& \sigma\left(pp\to
H^0A^0\right) \left[{2+x^2_{HA} \over 1+x^2_{HA}}
{ x_{HA}^2\over 1+ x_{HA}^2} \right]
 \mbox{BF}(H^0\to H^\pm W^{\mp^*})\nonumber\\
 &\times&\mbox{BF}(A^0\to H^\pm W^{\mp^*})
 \left[ \mbox{BF}(H^\pm\to H^{\pm\pm}W^{\mp^*})\right]^2
\left[\mbox{BF}(H^{\pm\pm}\to \ell^\pm\ell^\pm)\right]^2.
 \label{4l4W}
\end{eqnarray}
As expected, the cross-sections vanish for $x_{HA} \to 0$
recovering the lepton number conserving limit. In the limit of $x_{HA}\gg1$ (the maximal
lepton number violation), the  mixing factors become one and the SS4L numbers are controlled only by
the branching fractions of $H^0$ and $A^0$.

\begin{figure}[t]
 \begin{center}
 \includegraphics[width=45mm,angle=-90]{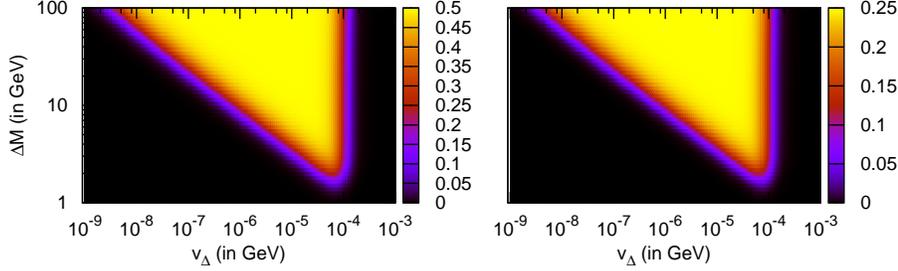}
 \end{center}
 \caption{Product of branching fractions for processes \ref{pro:1}
  (left), and \ref{pro:3} (right).
 \label{BFprod}}
 \end{figure}

\begin{figure}[t]
\begin{center}
 \includegraphics[width=80mm,angle=-90]{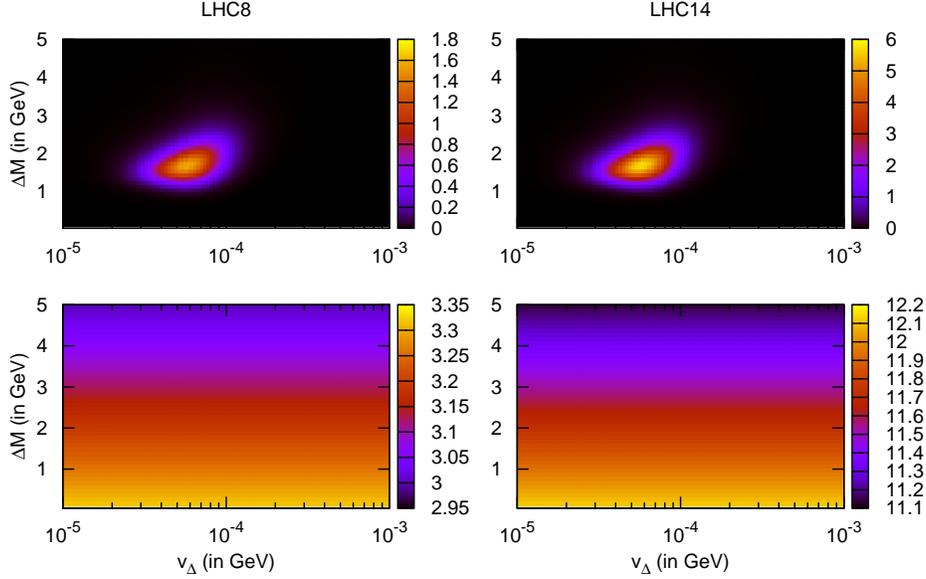}
\end{center}
\caption{Bottom panels show  cross-sections (in fb) for $H^\pm
H^0/H^\pm A^0/H^0A^0$ production, and top panels for same-sign
tetra-lepton events. Left (right) panels are for LHC8 (LHC14).
 \label{comb-cross}}
\end{figure}

The cross section for SS4L signal depend on $\vdel$ through decay branching fractions.
In Fig.~\ref{BFprod}, we show product of BFs which occur in the evaluation of cross-section for
processes \ref{pro:1}  (left figure) and for process \ref{pro:3}
(right figure) in $\Delta M-v_\Delta$ plane. One can see from
these figures that large parameter space are favourable. In Fig.~\ref{comb-cross}
(bottom), we show sum of cross-sections for processes \ref{pro:1}
and \ref{pro:3} which can finally give same-sign tetra-leptons.
The cross-sections are independent of $v_\Delta$ and steadily
decreases with the rise of $\Delta M$. In Fig.~\ref{comb-cross}
(top), we show cross-sections for
$\ell^\pm\ell^\pm\ell^\pm\ell^\pm$ signal at LHC8 and LHC14 in the
$\Delta M-v_\Delta$ plane. The Fig.~\ref{comb-cross} (top) is
obtained by superposing the Figs.~\ref{comb-cross} (bottom) and
\ref{BFprod} and multiplying with the oscillation factor as in
Eqs.~(\ref{4l3W},\ref{4l4W}). The doubly charged Higgs mass is
taken to be 400 GeV. 
One can see from
Fig.~\ref{comb-cross} that the same-sign tetra-lepton
cross-section is maximized for $v_\Delta = (10^{-4} - 10^{-5})$
GeV and $\Delta M=(1-2)$ GeV in accordance with the rough
estimate. 
If the parameter region happens to be near the limited bright region in Fig.~\ref{comb-cross},
one can find in addition for SS4L allowing to get information about $v_\Delta$ and $\Delta M$. 

%


Now we choose a benchmark point with $\vdel=7\times10^{-5}$ GeV,
$\Delta M=1.5$ GeV and $M_{H^\pm\pm}=400$ GeV which  gives $\delta
M_{HA} =3.68 \times 10^{-11}$ GeV, $\Gamma_{H^0/A^0} =3.73\times
10^{-11}$ GeV, and thus $x^2_{HA}/(1+x_{HA}^2) =0.79$.


So far, we have not distinguished among flavors of charged leptons
i.e.,  $e, \mu, \tau$ in our analysis. However, we know that at
the LHC, $\tau$ leptons are more difficult to identify.
$\tau$ leptons can decay leptonically $\tau\to
e\nu_e\bar \nu_\tau$ and $\tau\to \mu\nu_\mu\bar \nu_\tau$ with
branching fractions of 17\% each. These $e$'s and $\mu$'s are less
energetic than their parent $\tau$'s. On the other hand, decays
$H^{\pm,\pm}\to e^\pm e^\pm/\mu^\pm\mu^\pm/e^\pm\mu^\pm$ are much
cleaner and produce the two energetic $e$ and $\mu$ closer to
invariant mass $M_{H^{\pm\pm}}$. For the collider analysis
including lepton flavor dependence, we consider the full neutrino
mass matrices calculated for
the NH and IH, respectively. In the NH case, the BF of the
$H^{\pm\pm}$ decay to the $e$ and $\mu$ final states is only 32\%, and
thus $H^{\pm\pm}$ decay mainly to same-sign $\tau$ pairs. Of
course, the leptonic decays of these $\tau$'s to $e/\mu$ are
included in our analysis. On the other hand, for the IH case, the $H^{\pm\pm}$
decay BF to the $e$ and $\mu$ final states is 60\% and thus we
expect to have more SS4L signal events compared to the NH case.




The only potential background can come from multi $W$
productions demanding
four $W^\pm$ decaying leptonically. At the lowest order, cross-section is proportional to
$\alpha_{EW}^7$ and, at one loop, would be suppressed by
$\alpha_S^2\alpha_{EW}^8$ times loop-suppression factor. Thus, the
background for SS4L final states is practically zero at the LHC.

Since there is negligible background, the selection  criteria for
the leptons are very trivial. So, the basic cuts like $p_T>20$ GeV 
and $|\eta|<$2.5 for all leptons would be sufficient to detect our signal.
We use \texttt{CTEQ6L} parton  distribution
function (PDF) and the renormalization/factorization scale is set
at $2M_{H^+}$. We use \texttt{CALCHEP}  to
generate the parton level events for the relevant processes and 
\texttt{PYTHIA} for fragmentation and 
initial/final state radiations. 
%


The SS4L signal for $M_{H^{++}}=400$ GeV might
be barely observable at LHC8 for the IH case, but the event number
is too small to reconstruct its mass. But, at LHC14
with 100 fb$^{-1}$ of integrated luminosity there would be enough
number of events to look for the doubly charged Higgs mass of
$M_{H^{++}}=400$ GeV.
 Assuming the criteria of 10 signal events for the claim of discovery, we find that
$H^{\pm\pm}$ with mass $M_{H^{\pm\pm}}$ as large as  600 GeV and
700 GeV can be probed for NH and IH scenario respectively at the
LHC14 with 100 fb$^{-1}$ of integrated luminosity.

\begin{table}[h]
\begin{center}
\begin{tabular}{l|c|c}
\hline
                                &   Pre-selection   &   Selection\\
\hline
$\ell^\pm\ell^\pm\ell^\pm\ell^\pm ~~ \mbox{(LHC8-NH)}$  &   4      &   3  \\
$\ell^\pm\ell^\pm\ell^\pm\ell^\pm ~~ \mbox{(LHC8-IH)}$  &  9 & 8 \\
\hline
$\ell^\pm\ell^\pm\ell^\pm\ell^\pm ~ \mbox{(LHC14-NH)}$   &   110     &   94  \\
$\ell^\pm\ell^\pm\ell^\pm\ell^\pm ~ \mbox{(LHC14-IH)}$   &   240     &   210 \\
\hline
\end{tabular}\caption{\label{events}
Number  of events for SS4L signals before and after selection cuts for both NH and IH
scenarios at LHC8 and LHC14 with 15 fb$^{-1}$ and 100 fb$^{-1}$ of integrated luminosities respectively.}
\end{center}
\end{table}

\begin{figure}[h]
\begin{center}
\includegraphics[scale=0.5,angle=-90]{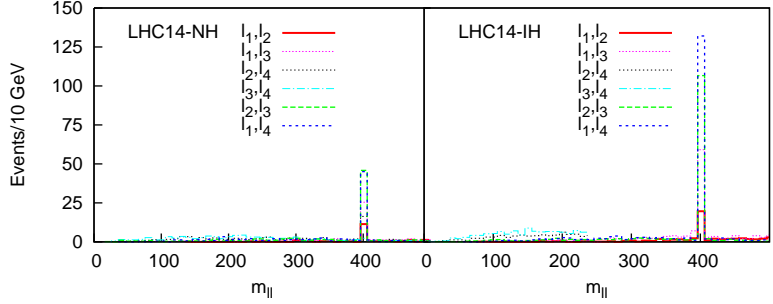}
\end{center}
\caption{Invariant mass from 4$\ell^\pm$ final states.\label{InvMass}}
\end{figure}

In Fig.~\ref{InvMass}, we plot the reconstructed  $H^{\pm\pm}$
mass from the sample of selected SS4L events for
both NH and IH neutrino mass scenarios at LHC14 with 100 fb$^{-1}$
integrated luminosity. The peaks in all plots correspond to
$H^{\pm\pm}\to ee/e\mu/\mu\mu$ decays while the broad part
(off-peak) of distribution correspond to $\tau$ decays. 

\section{Conclusion}

We point out that a remarkable signal of  SS4L,
 $l^\pm l^\pm l^\pm l^\pm$, is allowed in the type II seesaw mechanism.
Such a signal at the LHC
strongly depends on the mass splitting $\Delta M$ and
the triplet vev $v_{\Delta}$
When the doubly charged component $H^{\pm\pm}$ is the lightest,
larger $\Delta M$ allows more efficient gauge decay of the neutral component
to the singly charged one and then to the doubly charged  one. Thus, a pair production
of the triplet components at colliders can end up with producing $H^{\pm\pm}H^{\pm\pm}$
whose branching fraction to SS4L becomes larger for smaller $v_\Delta$.
Another crucial ingredient for increasing the SS4L signal number is the $H^0$-$A^0$ mixing
parameter $x_{HA}$ which becomes smaller for smaller $v_\Delta$ and larger $\Delta M$.
Therefore, there appear optimal values of the model parameters which maximize the same-sign
tetra-lepton signal.

After studying such a behavior in the $\Delta M-v_\Delta$ plane, we identified a benchmark point
with $\Delta M=1.5$ GeV and $v_\Delta = 7\times10^{-5}$ GeV for
maximized the signal numbers at the LHC.
After making the typical selection cuts to identify four same-sign leptons, we have shown that
one can obtain sizable event numbers for $M_{H^{\pm\pm}} =400$  GeV with integrated luminosity
of 100 fb$^{-1}$ at the LHC14 to reconstruct the doubly charged Higgs mass
in both cases of the normal and inverted neutrino mass hierarchy.

\end{document}